\documentclass[reprint,aps,pre,amsmath,amssymb,superscriptaddress,showpacs]{revtex4-1}
\usepackage{graphicx}

\begin{document}

\title{Complex hypergraphs}

\author{Alexei Vazquez}
\email{alexei@nodeslinks.com}
\affiliation{Nodes \& Links Ltd, Salisbury House, Station Road, Cambridge, CB1 2LA, UK}


\begin{abstract}
Providing an abstract representation of natural and human complex structures is a challenging problem. Accounting for the system heterogenous components while allowing for analytical tractability is a difficult balance. Here I introduce complex hypergraphs (chygraphs), bringing together concepts from hypergraphs, multi-layer networks, simplicial complexes and hyperstructures. To illustrate the applicability of this combinatorial structure I calculate the component sizes statistics and identify the transition to a giant component. To this end I introduce a vectorization technique that tackles the multi-level nature of chygraphs. I conclude that chygraphs are a unifying representation of complex systems allowing for analytical insight.
\end{abstract}

\maketitle

\section{Introduction}

Graphs are structural abstractions of many-particle systems, with vertices representing particles and edges the pairwise interactions between them \cite{albert02}. Hypergraphs are an extension allowing for interactions between two or more vertices \cite{ghoshal09, vazquez09hg, coutinho20, civilini21, sun21}. Multiplex networks introduce layers accounting for different interaction types \cite{gomez13}.
Simplicial complexes extend connectivity to hierarchical structures of inclusion \cite{petri18, courtney18, iacopini19}.

There are self-referential constructions as well. Joslyn and Nowak introduced ubergraphs \cite{joslyn17}, a combinatorial structure where edges can contain other edges. ubergraphs has been used to organize relations in knowledge databases \cite{yadati21}. At a higher level of abstraction, Baas introduced higher order structures, also called hyperstructures, using concepts from category theory \cite{baas19}.

In practice what is a suitable representation is a matter of balance. We would like the flexibility of hypergraphs to go beyond pairwise interactions, the possibility of multiple layers and the hierarchical inclusion structure of simplicial complexes and hyperstructures. In turn we would like the simplicity of graphs to calculate aggregate properties, such as the size of the giant component. 

Here I introduce complex hypergraphs as a flexible combinatorial structure. General enough to include a big volume of previous work and new applications. Simple enough to allow for analytical calculations. The work is organized as follows. In Sect. \ref{definition} I define complex hypergraphs and comment on its relation with other structures. In Sects. \ref{percolation1} and \ref{percolation2} I provide analytical methods to characterize the emergence of the chygraph giant component, using a generating function formalism. I discuss particular cases and provide numerical examples validating the analytical results. I finish with some key conclusions in Sec.\ref{conclusions}.

\section{Key definitions}
\label{definition}

\subsection{Motivation}

Complex systems are simplified to facilitate their analysis. Gradually we remove some of the simplifications and move closer to the real system. For example, the system of scientific publications has been represented by different network structures depending on the question asked \cite{newman01, barabasi02}. Citation networks indicate the flow of knowledge along publications \cite{redner98, lehmann03}. In citation networks nodes are publications and citations are represented by directed links. Co-authorship networks are better suited when focusing on collaborations. Authors-publications networks can be further expanded to explicitly represent authors and publications, resulting in the bipartite graph \cite{newman01}. The same authors-publications relations can be represented as an authors hypergraph, where publications are hyperedges associating one, two or more authors \cite{vasilyeva21}. These networks and hypergraphs are simplifications loosing some aspects of the original system. We need a more complete representation with a richer combinatorial structure. A scientific publication contains both authors and references. The document internal structure can be represented by a hypergraph with two edges, the list of authors and references. On top of that, the publication is a vertex in a higher order structure where the building blocks are authors and publications. Informally speaking, a hypergraph of hypergraphs, a complex hypergraph.

\subsection{Complex systems}

Here I use the term complex in the structural sense: made of different parts. I make a distinction between the parts that are not decomposable into other parts, the atoms, and the complexes that are made of other parts, including atoms. The atoms could have a finer structure, but they have been chosen as the primary building blocks. These preliminaries lead to the self-consistent definition of complex system.

{\em Definition. A complex system} is a set of atoms and complexes, where complexes are made of atoms and other complexes.

The latter serves as a philosophical definition of complex system. For practical applications we need a precise mathematical construction. The choice depends on the internal structure of the complex.

\subsection{Hypergraphs}

A hypergraph ${\cal H}(V,E)$ is a vertex set $V$ and a hyperedge set $E$, where a hyperedge is a subset of $V$. Just for the sake of illustration, I rewrite this definition based on the wording of complex systems introduced above. A hypergraph ${\cal H}(A,C)$ is an atom set $A$ and a complex set $C$, where the complexes are subsets (hyperedges) of $A$. Obviously hypergraphs do not posses the self-referential property of complex systems. They are simple structures. 

\subsection{ubergraphs}

To make hypergraphs self-referential, Joslyn and Nowak introduced ubergraphs \cite{joslyn17}, a combinatorial structure where hyperedges can contain other hyperedges. Here I redefine them using the wording of complex systems introduced above.

{\em Definition. An ubergraph} ${\cal U}(A,C)$ is a set of atoms $A$ and a set of complexes $C$, where the complexes are subsets (hyperedges) of $A\cup C$.

\begin{figure}[t]
\includegraphics[width=3.3in]{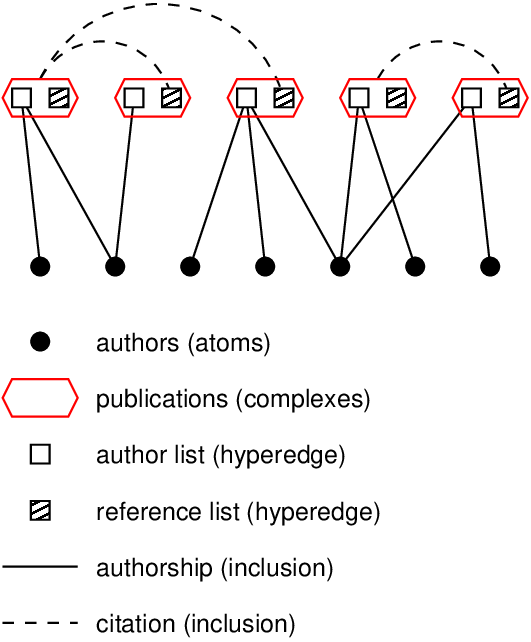}
\caption{A chygraph example. This chygraph is a representation of the system of scientific publications, with atoms representing authors, complexes publications and inclusions citations.}
\label{fig_publications}
\end{figure}

\subsection{Complex hypergraphs}

To add more internal structure I choose hypergraphs. That leads to the definition of complex hypergraphs.

{\em Definition. A complex hypergraph (chygraph)} $\chi (A,C)$ is a set of atoms $A$ and a set of complexes $C$, where the complexes are hypergraphs with vertex sets in $A\cup C$.

Let us unravel this definition with some examples. A graph ${\cal G}(V,E)$ is represented by the chygraph $\chi(V, E)$ where the complexes are edges. A multiplex graph \cite{gomez13} with layer graphs $G_l(V,E_l)$, $l=1,\ldots,L$, is represented by the chygraph $\chi( V, \cup_{l=1}^LE_l)$ plus some partition structure discussed below. The system of scientific publications is represented by the chygraph $\chi( A, \{ {\cal H}_i(A_i\cup R_i, \{A_i, R_i\})\})$, where atoms are authors, complexes are publications, the publications are represented by a hypergraph with two edges ($A_i$ for the authors and $R_i$ for the references) and the index $i$ runs across all publications. The two edges have no overlap and therefore the complexes representing scientific publications have two internal components (Fig. \ref{fig_publications}).

We can think of chygraphs as ubergraphs with two edge types: the edges in the complexes associated hypegraphs and the complexes. The complex act as a special edge that we designate as the basic unit of inclusion. In turn, ubergraphs are chygraphs where all intra-complex hypergraphs contain a single edge. Obviously in this case the intra-complex hypergraph has only one component with a size equal to the hyperedge cardinality. Therefore, all results derived in the following sections are valid for ubergraphs, after replacing component size by hyperedge cardinality. 

\subsection{Chygraphs properties}

Many properties of graphs/hypergraphs are carried on to chygraphs. To make a distinction from the metrics associated with the complexes  hypergraph structure, I will use greek letters to name quantities at the chygraph level. Given a complex $j$, let ${\cal H}(V_j,E_j)$ be its associated hypergraph. Let $i\in A\cup C$ be an atom or complex, where $i=1,\ldots,n$ and $n=| A\cup C|$. I will use the notation $i\in C_j$ as equivalent to $i\in V_j$. The chy-adjacency matrix $\alpha$ is the $n\times n$ matrix with matrix elements
\begin{equation}
\alpha_{ij} = \left\{\begin{array}{ll}
1 & {\rm if}\ i\in C_j\ ,\\
0 & {\rm otherwise}\ ,
\end{array}\right.
\label{a}
\end{equation}
where $i,j\in A\cup C$. The associated vertex chy-degrees
 \begin{equation}
 \kappa_i = \sum_j \alpha_{ij}\ .
\label{k}
\end{equation}
Additional metrics are needed to characterize potential multi-level structure. Chygraphs may contain multiple levels of inclusion. At the lower level we have atoms, whose fine structure is null or not specified. One level above we have complexes, their internal structure being specified as hypergraph containing atoms and/or other complexes as vertices. In some systems it makes sense to define higher levels of inclusions. For example, when describing human populations by location, we speak of neighborhoods, cities, countries, continents and the world. This inclusion hierarchy leads to the definition of chygraph length. 

{\em Definition: chygraph length $L(\chi)$}. Let $\chi(A,C=\{{\cal H}_i(V_i, E_i)\})$ be a chygraph. Let  $\Pi=\Pi_1\cup\ldots\cup\Pi_l$ be a partition of $C$ that is non-intercepting ($\Pi_i\cap \Pi_j=\emptyset$ for $i\neq j$), hierarchical (if $C_i\in \Pi_j$ then $V_i\subset A\cup(\cup_{k\leq j}\Pi_k)$) and complexes within the same partition have similar statistical properties. The chygraph length, denoted by $L(\chi)$, is the maximum $l$ among all such partitions.

The differentiation of partitions by statistical properties allows for the specification of multi-type structures. This is the case in multiplex graphs and hypergraphs. In this context two layers may have the same vertex set but the complexes may have different statistical properties depending on the layer.

As in the case of higher order networks \cite{dedomenico13}, we could extend other metrics such as the clustering coefficient and the entropy to chygraphs. Here I have limited my attention to metrics that are needed in the following sections. 

\section{Percolation theory I: no intra-layer inclusions}
\label{percolation1}

In spite of its complexity the chygraph combinatorial structure is suitable for analytical treatment. I focus on percolation, a central problem in graph theory. 

The generating functions technique has been used to solve percolation problems in graphs and hypergraphs \cite{callaway00, ghoshal09, coutinho20, sun21}. It follows a simple recipe: express the generating function of the component sizes distribution as a recursive function of itself, modulated by the generating functions of other relevant distributions. In the context of chygraphs, the key quantities are the component sizes $\sigma^l$ and excess component sizes $\bar{\sigma}^{ml}$ when the components are sampled from layer $l$ at random or coming from another layer $m$, respectively. Here we need to be precise about how we arrive from another layer.

We can arrive to a reference complex in two different ways. From the complexes it includes (coming from below)  or from the complexes where it is included (coming from above). If there are no intra-layer connections the notation $\bar{\sigma}^{ml}$ does not require any further specification. When $m<l$ it is clear we are arriving from below. In contrast, when $m>l$ we are arriving from above. In this section I assume there are no intra-layer connections. The case with intra-layer connections will be the subject of next section.

The component sizes depend on the joint distribution of chy-degrees $\Hat{\kappa}_l$, in-complex hypergraph component sizes $\Check{s}^l$ and their excess equivalents $\Check{\bar{\kappa}}^m_l$ and $\Check{\bar{s}}^m_l$ when reached from layer $m$. The notation $\Check{s}_l = (s_{l0}, s_{l1}\ldots, s_{ll-1}$) indicates that a component within a complex at layer $l$ is composed of vertices from $A,\Pi_1,\ldots,\Pi_{l-1}$.  In turn, $\Hat{\kappa}_l=(\kappa_{ll+1},\ldots,\kappa_{lL})$ indicates that the chy-degree of a vertex at layer $l$ is decomposed into chy-degrees to vertices in layers $\Pi_{l+1}, \ldots, \Pi_L$. The probability generating functions of $\Hat{\kappa}_l$, $\Check{\bar{\kappa}}_{ml}$, $\Check{s}^l$, $\Check{\bar{s}}^{ml}$, $\sigma^l$ and $\bar{\sigma}^{ml}$ are denoted by $\Phi^l(\Hat{x}_l)$, $\Psi^{ml}(\Hat{x}_l)$, $G^l(\Check{x}_l)$, $U^{ml}(\Check{x}_l)$, $\Gamma^l(x)$ and $\Upsilon^{ml}(x)$, respectively.  Since they are generating functions of probability distributions, they are all equal to 1 when evaluated at $x=1$ and their first derivatives are equal to the corresponding expected values.

\subsection{Mean component size}

The definition of chygraph is translated into a set of self-consistent equations for the component size generating functions. As a guideline, we arrive to a reference complex and from the complex we navigate the chygraph.  Following the chygraph adjacency matrix, leading us to complexes including the reference complex, or going through the complexes included in the reference complex hypergraph. More precisely,
\begin{eqnarray}
\Gamma^l(x) &=& x\Phi^l[ \Upsilon^{ll+1}(x), \ldots, \Upsilon^{lL}(x) ]  \nonumber\\
&\times& G^l[ \Upsilon^{l0}(x), \ldots, \Upsilon^{ll-1}(x) ],
\label{Gammal}
\end{eqnarray}
\begin{eqnarray}
\Upsilon^{ml}(x) &=& x\Psi^{ml}[ \Upsilon^{ll+1}(x), \ldots, \Upsilon^{lL}(x) ] \nonumber\\
&\times& U^{ml}[ \Upsilon^{l0}(x), \ldots, \Upsilon^{ll-1}(x) ],
\label{Upsilonl}
\end{eqnarray}
where $l=0,\ldots,L(\chi)$. Note that the first $x$ in the right hand side of these equations corresponds to the reference complex, the terms containing $\Phi^l[\cdots]$ and $\Psi^{ml}[\cdots]$ to navigations using the chygraph adjacency matrix (complexes including the reference complex) and the terms containing  $G^l[\cdots]$ and  $U^{ml}[\cdots]$ to navigations through the complexes in the reference complex hypergraph.

\subsection{Mean component size}

The mean excess component sizes $\langle\bar{\sigma}\rangle^{ml} = \dot{\Upsilon}^{ml}(1)$  can be calculated from Eq. (\ref{Upsilonl}), resulting in
\begin{equation}
\langle\bar{\sigma}\rangle^{ml} = 1 + 
\sum_{k=l+1}^L\langle\bar{\kappa}\rangle^m_{lk}\langle\bar{\sigma}\rangle^{lk}
+ \sum_{k=0}^{l-1}\ \langle \bar{s}\rangle^{m}_{lk} \langle\bar{\sigma}\rangle^{lk},
\label{barsigmal}
\end{equation}
where $l,m=0,\ldots,L$. Note that  $\langle\bar{\kappa}\rangle^m_{lk}\neq\langle\kappa\rangle_{lk}$ and $\langle \bar{s}\rangle^{m}_{lk}\neq \langle s\rangle_{lk}$ only when $k=m$. When we come from a layer $m$ into a layer $l$ and then return to layer $m$.  Now comes the vectorization trick.

The matrix equation (\ref{barsigmal}) can be solved by the vectorization method for matrix equations \cite{horn94}. The vectorization operator ${\rm vec}X$ transform a $(M,N)$ matrix into a $M\times N$ column vector by  stacking the columns of $X$. For example,
\begin{equation}
{\rm vec}X^{(2,2)} = \begin{bmatrix}
X^{00}\\ X^{10}\\ X^{01}\\X^{11}.
\end{bmatrix}
\end{equation}
To handle tensors with four indexes I generalize the vectorization operator. The vectorization operator acting on the $X^{(M, N)}_{(O, P)}$ tensor creates a $(M\times N, O\times P)$ matrix by stacking the upper indexes along columns and lower indexes along rows. For example,
\begin{equation}
{\rm vec}X^{(2,2)}_{(2,2)} = \begin{bmatrix}
X^{00}_{00} & X^{00}_{01} & X^{00}_{10} & X^{00}_{11}\\
X^{01}_{00} & X^{01}_{01} & X^{01}_{10} & X^{01}_{11}\\
X^{10}_{00} & X^{10}_{01} & X^{10}_{10} & X^{10}_{11}\\
X^{11}_{00} & X^{11}_{01} & X^{11}_{10} & X^{11}_{11}
\end{bmatrix}
\end{equation}
Applying vectorization Eq. (\ref{barsigmal}) is transformed to
\begin{equation}
{\rm vec}\{A\} {\rm vec}\{\langle\bar{\sigma}\rangle\} = {\rm vec}\{B\},
\label{vec_sigmal}
\end{equation}
where
\begin{equation}
A^{ml}_{nk} = \delta_{mn}\delta_{lk} 
- \langle\bar{\kappa}\rangle^{m}_{nk} \Theta_{k-n} \delta_{nl} 
- \langle\bar{s}\rangle^{m}_{nk} \Theta_{n-k} \delta_{nl},
\label{A}
\end{equation}
\begin{equation}
B^{ml} = 1,
\label{D}
\end{equation}
and $\Theta_i$ is an integer Heaviside step function ($\Theta_i=1$ if $i>0$ or 0 otherwise).
The linear systems of equations (\ref{vec_sigmal}) has Cramer's rule as a formal solution
\begin{equation}
{\rm vec}\{\langle\bar{\sigma}\rangle\}_i = \frac{\det({\rm vec}\{A\}_i)}{\det({\rm vec}\{A\})},
\label{sigma_cramer}
\end{equation}
where ${\rm vec}\{A\}_i$ is derived from ${\rm vec}\{A\}$ by replacing the $i$th column by ${\rm vec}\{B\}$. This solution is valid provided that $A$ is not singular. When $\det(A)\rightarrow0^+$ the mean component sizes diverge and the system achieve percolation. Therefore, the {\em chygraph critical percolation condition} is given by
\begin{equation}
\det({\rm vec}\{A\}) = 0.
\label{critical}
\end{equation}

\subsection{Giant component}

The equation for the mean component sizes (\ref{vec_sigmal}) is valid provided $\det({\rm vec}\{A\})>0$. In the following I demonstrate that $\det({\rm vec}\{A\})>0$ corresponds with the subcritical phase. Let $P^l$ be the probability that a vertex from layer $l$ selected at random does not belong to the giant component and let $Q^{ml}$ be the probability that a vertex at layer $l$ selected from a complex at layer $m$ does not belong to the giant component. These probabilities satisfy the self-consistent equations
\begin{equation}
P^l = \Phi^l[Q^{ll+1}, \ldots, Q^{lL}]\ G^l[ Q^{0l}, \ldots, Q^{l-1l} ],
\label{P}
\end{equation}
\begin{equation}
Q^{ml} = \Psi^{ml}[Q^{ll+1}, \ldots, Q^{lL}]\ U^{ml}[ Q^{0l}, \ldots, Q^{l-1l} ].
\label{Q}
\end{equation}
This system of equations does not have an explicit analytic solution. A solution can be found by successive approximations, where the left hand side is interpreted as the $t+1$ iteration after plugging in iteration $t$ into the right hand side. In particular, in the absence of a giant component, Eqs. (\ref{P}) and (\ref{Q}) admit the solution $P^l=Q^{ml}=1$. Let us assume that  $Q^{lm}=1-x^{ml}$, where $x^{ml}\rightarrow 0$. Keeping terms up to first order in $x^{ml}$ in Eq. (\ref{Q}) results in the recursive approximation equations
\begin{equation}
{\rm vec}\{x\}(t+1) = (I-{\rm vec}\{A\}) {\rm vec}\{x\}(t).
\label{xx}
\end{equation}
The linear map (\ref{xx}) converges to ${\rm vec}\{x\}=0$ if and only if $\Lambda({\rm vec}\{A\})>0$, where $\Lambda({\rm vec}\{A\})$ is the largest eigenvalue of ${\rm vec}\{A\}$. Therefore $\Lambda({\rm vec}\{A\})$ is the control parameter for the existence of a giant component. In the subcritical (supercritical) phase $\Lambda>0$ ($\Lambda<0$) there is not (there is) a giant component  and $P^l=1$ ($P^l<1$). The percolation transition takes place at the criticality condition $\Lambda({\rm vec}\{A\})=0$, which is equivalent to the Eq. (\ref{critical}).

\subsection{$L=1$}

When there is only one complex layer ($L(\chi)=1$) then ${\rm vec}\{A\}$ in Eq. (\ref{A}) is reduced to
\begin{equation}
{\rm vec}\{A\} = \begin{bmatrix}
1 & 0 & 0 & 0\\
0 & 1 & A^{01}_{10} & 0\\
0 & A^{10}_{01} & 1 & 0\\
0 & 0 & 0 & 1\end{bmatrix}
\label{AL1}
\end{equation}
with determinant
\begin{equation}
\det({\rm vec}\{A\}) = 1 - A^{10}_{01} A^{01}_{10} = 1 - \bar{k}^1_{01}\bar{s}^{0}_{10}.
\label{AL1det}
\end{equation}

A hypergraph ${\cal H}(V,E)$ is mapped to a one layer chygraph where the complexes are the hypergraph edges: $\chi(V,\{{\cal H}(V,e_l), e_l\in E\})$. In this case the excess component sizes $\bar{s}^{0}_{10}$ is the excess hyperedges cardinality $\langle\bar{c}\rangle$ and the excess atoms degree $\bar{k}^1_{01}$ is the vertices excess degree $\langle\bar{k}\rangle$. Substituting into Eq. (\ref{AL1det}) one obtains the critical condition for hypergraphs: $\langle\bar{c}\rangle\langle\bar{k}\rangle = 1$, in agreement with the result of Coutinho {\em et al} \cite{coutinho20}. Furthermore, graphs are hypergraphs with excess cardinality 1 and, therefore, the criticality condition reduces to $\langle\bar{k}\rangle = 1$, as previously reported by Molloy and Reed \cite{molloy98} and Callaway {\em et al} \cite{callaway00}.

\subsection{$L=2$}

When there are two complex layers ($L(\chi)=2$) then ${\rm vec}\{A\}$ in Eq. (\ref{A}) is reduced to
\begin{equation}
{\rm vec}\{A\} = \begin{bmatrix}
1 & 0 & 0 & 0 & 0 & 0 & 0 & 0 & 0\\
0 & 1 & 0 & A^{01}_{10} & 0 & A^{01}_{12} & 0 & 0 & 0\\
0 & 0 & 1 & 0 & 0 & 0 & A^{02}_{20} & A^{02}_{21} & 0\\
0 & A^{10}_{01} & A^{10}_{02} & 1 & 0 & 0 & 0 & 0\\
0 & 0 & 0 & 0 & 1 & 0 & 0 & 0 & 0\\
0 & 0 & 0 & 0 & 0 & 1 & A^{12}_{20} & A^{12}_{21} & 0\\
0 & A^{20}_{01} & A^{20}_{02} & 0 & 0 & 1 & 0 & 0\\
0 & 0 & 0 & A^{21}_{10} & 0 & A^{21}_{12} & 0 & 1 & 0 \\
0 & 0 & 0 & 0 & 0 & 0 & 0 & 0 & 1\end{bmatrix},
\label{AL2}
\end{equation}
with determinant
\begin{equation}
\det({\rm vec}\{A\}) = \begin{vmatrix}
1 & 0 & A^{01}_{10} & A^{01}_{12} & 0 & 0\\
0 & 1 & 0 & 0 & A^{02}_{20} & A^{02}_{21}\\
A^{10}_{01} & A^{10}_{02} & 1 & 0 & 0 & 0 \\
0 & 0 & 0 & 1 & A^{12}_{20} & A^{12}_{21}\\
A^{20}_{01} & A^{20}_{02} & 0 & 0 & 1 & 0\\
0 & 0 & A^{21}_{10} & A^{21}_{12} & 0 & 1\end{vmatrix}.
\label{AL2det}
\end{equation}
We could expand the determinant but it would lead to a complicated algebraic expression. For most practical purposes it is best to work with the explicit matrix form in Eq. (\ref{AL2det}), including numerical calculations. Further simplifications are obtained when we consider combinatorial constructions with additional constraints.

\subsection{Hierarchical inclusion}

There are many systems where inclusion follows a hierarchy. Geographical zooming for example \cite{okabe96}. It is straightforward to build hierarchical inclusion in chygraphs: atoms included in layer 1 complexes, layer 1 complexes included in layer 2 complexes, ... , layer $L-1$ complexes included in layer $L$ complexes. In this context the tensor elements of $A^{ml}_{nk}$ are zero if $|m-l|\neq 1$ or $|n-k|\neq1$
For the case $L=2$ this constraint leads to the elimination of the second and fifth row and the second and fifth column in Eq. (\ref{AL2det}), resulting in
\begin{equation}
\det({\rm vec}\{A\}) = \begin{vmatrix}
1 & A^{01}_{10} & A^{01}_{12} & 0\\
A^{10}_{01} & 1 & 0 & 0\\
0 & 0 & 1 & A^{12}_{21}\\
0 & A^{21}_{10} & A^{21}_{12} & 1\end{vmatrix}.
\label{AL2det_hierachy}
\end{equation}
Expanding the latter determinant and substituting the explicit form of the tensor $A$ (Eq. (\ref{A})) we obtain
\begin{eqnarray}
\det({\rm vec}\{A\}) &=& (1 - \langle\bar{\kappa}\rangle^1_{01} \langle\bar{s}\rangle^0_{10} )
(1 - \langle\bar{\kappa}\rangle^2_{12} \langle\bar{s}\rangle^1_{21} ) \nonumber\\
&-& \langle\bar{\kappa}\rangle^0_{12} \langle\bar{s}\rangle^2_{10} \langle\bar{\kappa}\rangle^1_{01} \langle\bar{s}\rangle^1_{21}
\label{AL2det_hierarchy_expanded}
\end{eqnarray}
Within the parentheses we have the criticality condition for each layer alone. The last term represents the interactions including both layers. The latter is the bona fide complexity of hierarchical inclusion. The terms within the first two parenthesis can be positive, meaning no standard 1 layer percolation, and $\det({\rm vec}\{A\})$ can become 0 due to the last interaction term. 

\subsection{Multiplex hypergraphs}

Multiplex hypergraphs is another type of multi-layer structure \cite{sun21}. A multiplex hypergraph is a set of hypergraphs $\{{\cal H}_l(V,E_l), l=1,\ldots,L\}$ with the same set of vertices. Note that when the statistical properties of the hypergraphs ${\cal H}_l$ are different the multiplex hypergraph is not statistically equivalent to ${\cal H}(V,\cup_{l=1}^LE_l)$. A multiplex hypergraph can be mapped to the chygraph $\chi(V,\cup_{l=1}^LE_l, \Pi=E_1\ldots,E_L)$, where all edges are represented by complexes and the complexes are partitioned according to the hypergraph they originated from. For the case $L=2$, the absence of inclusion of complexes into complexes leads to the elimination of the fourth and sixth row and the fourth and sixth column of (\ref{AL2det}), resulting in
\begin{equation}
\det({\rm vec}\{A\}) = \begin{vmatrix}
1 & 0 & A^{01}_{10} & 0\\
0 & 1 & 0 & A^{02}_{20}\\
A^{10}_{01} & A^{10}_{02} & 1 & 0\\
A^{20}_{01} & A^{20}_{02} & 0 & 1\end{vmatrix}.
\label{AL2det_multiplex}
\end{equation}
Expanding the latter determinant and substituting the explicit form of the tensor $A$ (Eq. (\ref{A})) we obtain
\begin{eqnarray}
\det({\rm vec}\{A\}) &=& (1 - \bar{k}^1_{01}\bar{s}^{0}_{10})(1 - \bar{k}^2_{02}\bar{s}^{0}_{20}) \nonumber\\
&-& \bar{k}^2_{01} \bar{s}^{0}_{10} \bar{k}^1_{02} \bar{s}^{0}_{20}.
\label{AL2det_multiplex_expanded_explicit}
\end{eqnarray}
Within the first two parentheses there is the contribution of each hypergraph layer alone. The last term represents the interaction between the two hypergraphs via the vertices. The latter is the bona fide complexity of multiplex hypergraphs. The terms within the first two parenthesis can be positive, meaning no standard hypergraph percolation, and $\det({\rm vec}\{A\})$ can become 0 due to the last interaction term.

\begin{figure}[t]
\includegraphics[width=3.3in]{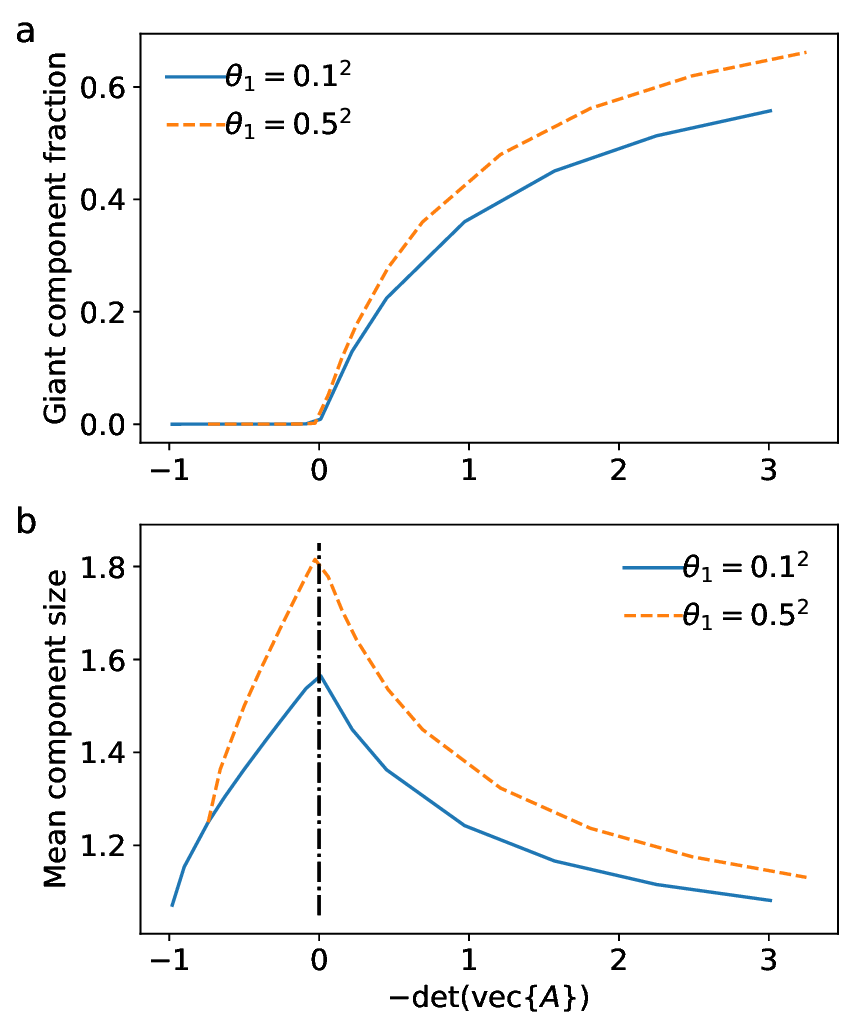}
\caption{{\bf Without intra-layer inclusions.} Numerical estimation of (A) the giant component fraction and (B) the mean component size as a function of $-\det({\rm vec}\{A\})$ for a random chygraph with a multiplex structure: one layer of $n_0$ atoms, two layers of complexes with $n_1$ and $n_2$ complexes and $m_i$ inclusions of a randomly chosen atom into a randomly chosen complex in layer $i$, $i=1,2$. The model parameters are $n_0=n_1=n_2=10^6$, two values of $\theta_1=m_1^2/(n_0n_1)$ indicated in the legend and $\theta_2=m_2^2/(n_0n_2)$ in the interval $[0.01,4]$. Average was taken over 10 random chygraphs.}
\label{fig_random_chygraph1}
\end{figure}

\subsection{Numerical example}

To test the analytical results I will use a random chygraph model with a multiplex structure: one layer of $n_0$ atoms, two layers of complexes with $n_1$ and $n_2$ complexes and $m_i$ inclusions of a randomly chosen atom into a randomly chosen complex in layer $i$, $i=1,2$. Since inclusions are random both the chydegrees and the component sizes have Poisson distributions with averages $k_{0i}=m_i/n_i$ and $s_{i0}=m_i/n_0$, $i=1,2$. For the Poisson distribution the excess average coincides with the average and therefore $\bar{k}^j_{0i}=m_i/n_i$ and $\bar{s}^j_{i0}=m_i/n_0$, $i=1,2$. Substituting these values into Eq. (\ref{AL2det_multiplex_expanded_explicit}) we arrive to the criticality condition
\begin{equation}
\det({\rm vec}\{A\}) = (1-\theta_1)(1-\theta_2) -\theta_1\theta_2
\label{AL2det_multiplex_expanded_explicit_random}
\end{equation}
where $\theta_i=m_i^2/n_in_0$, $i=1,2$.

To test this expression I computed the giant component fraction and the mean component size excluding the giant component numerically. To this end I project the chygraph into a network where nodes represent atoms and complexes, while links represent inclusion relations. Figure \ref{fig_random_chygraph1} reports the numerical estimation of the giant component fraction and the mean component size as a function of $-\det({\rm vec}\{A\})$, computed using Eq. (\ref{AL2det_multiplex_expanded_explicit_random}). There is a phase transition at $-\det({\rm vec}\{A\})=0$, with the emergence of a giant component and a maximum of the mean component size excluding the giant component. The agreement demonstrates the validity of the analytical results for multiplex chygraphs.

In turns out this example is a validation for the hierarchical inclusion as well. Consider another random chygraph model with the following inclusion structure: one layer of $n_0$ atoms, a layer labeled 1 with $n_1$ complexes and $m_1$ inclusions of a randomly chosen atom into a randomly chosen complex from layer 1, and a layer labeled 2  with $n_2$ complexes and $m_2$ inclusions of a randomly chosen complex from layer 1 into a randomly chosen complex from layer 2. One can check that the criticality condition for chygraphs with hierarchical inclusion in Eq. (\ref{AL2det_hierarchy_expanded}) is reduced to Eq. (\ref{AL2det_multiplex_expanded_explicit_random}). Furthermore,  projecting the chygraph into a network, where nodes represent atoms and complexes,  while links represent inclusion relations, we obtain the same network as for the multiplex chygraph described above. Note that this equivalence holds for $L=2$. For $L>2$ there is no equivalence between random chygraphs with a multiplex or hierarchical inclusion structure. For the multiplex structure the network projection has one layer connected to $L-1$ layers. For the hierarchical inclusion the network projection has $L$ ordered layers with links between adjacent ayers only.

\section{Percolation theory II: with intra-layer inclusions}
\label{percolation2}

We can arrive to a reference complex from the complexes it includes (from below) or where it is included (from above). When there are no intra-layer connections the pair of indexes $ml$ is sufficient to distinguish how we arrived to the reference complex. If $m<l$ we came from below. If $m>l$ we cam from above. However, when there are intra-layer inclusions the two indexes are not sufficient to specify how we arrived to a complex for the case $m=l$. We need an extra index. In the context of the generating function formalism that means we need two component size generating functions when arriving to a complex from an atom or another complex. I will denote them by $\Upsilon^{ml}_{-}$ when arriving from below and $\Upsilon^{ml}_{+}$ when arriving from above. With this distinction Eq. (\ref{Gammal}) and (\ref{Upsilonl}) are rewritten as
\begin{eqnarray}
\Gamma^l(x) &=& x\Phi^l[ \Upsilon^{ll}_{-}(x), \ldots, \Upsilon^{lL}_{-}(x) ]  \nonumber\\
&\times& G^l[ \Upsilon^{l0}_{+}(x), \ldots, \Upsilon_{+}^{ll}(x) ],
\label{Gammal2}
\end{eqnarray}
\begin{eqnarray}
\Upsilon^{ml}_{i}(x) &=& x\Psi^{ml}_{i}[ \Upsilon^{ll}_{-}(x), \ldots, \Upsilon^{lL}_{-}(x) ] \nonumber\\
&\times& U^{ml}_{i}[ \Upsilon_{+}^{l0}(x), \ldots, \Upsilon_{+}^{ll}(x) ],
\label{Upsilonl2}
\end{eqnarray}
where $l=0,\ldots,L(\chi)$ and
\begin{equation}
\Psi^{ml}_{i}(x) = \left\{ \begin{array}{ll}
\Phi^l(x) & {\rm for}\ i=-,\\
\Psi^{ml}(x) & {\rm for}\ i=+,
\end{array}\right.
\label{psii}
\end{equation}
\begin{equation}
U^{ml}_{i}(x) = \left\{ \begin{array}{ll}
G^l(x) & {\rm for}\ i=-,\\
U^{ml}(x) & {\rm for}\ i=+.
\end{array}\right.
\label{psii}
\end{equation}
Note I have implicitly assumed that the chydegrees and the intra-complex hypergraph components are independent.

The mean excess component sizes $\langle\bar{\sigma}\rangle^{ml}_{i} = \dot{\Upsilon}^{ml}_{i}(1)$, $i=-,+$,  can be calculated from Eq. (\ref{Upsilonl2}), resulting in
\begin{equation}
\langle\bar{\sigma}\rangle^{ml}_{i} =  1 + \sum_{k=l}^L\langle\bar{\kappa}\rangle^m_{ilk} \langle\bar{\sigma}\rangle^{lk}_{-} + \sum_{k=0}^{l}\ \langle \bar{s}\rangle^{m}_{ilk} \langle\bar{\sigma}\rangle^{lk}_{+},
\label{barsigmal2}
\end{equation}
where
\begin{equation}
\langle\bar{\kappa}\rangle^m_{ilk} = \left\{ \begin{array}{ll}
\langle\kappa\rangle_{lk} & {\rm for}\ i=-,\\
\langle\bar{\kappa}\rangle^m_{lk} & {\rm for}\ i=+,
\end{array}\right.
\label{psii}
\end{equation}
\begin{equation}
\langle \bar{s}\rangle^{m}_{ilk} = \left\{ \begin{array}{ll}
\langle \bar{s}\rangle^{m}_{lk} & {\rm for}\ i=-,\\
\langle s\rangle_{lk} & {\rm for}\ i=+.
\end{array}\right.
\label{psii}
\end{equation}

We can apply the vectorization trick to transform Eq. (\ref{barsigmal2}) into a standard linear system of equations. To do so we need to apply the vectorization trick twice. Once as done before for the indexes $(m,l,k)$ and another one for the index $i$. After the first vectorization Eq. (\ref{barsigmal2}) can be rewritten as
\begin{equation}
\begin{bmatrix}
{\rm vec}\{A_{--}\} & {\rm vec}\{A_{-+}\}\\
{\rm vec}\{A_{+-}\} & {\rm vec}\{A_{++}\}
\end{bmatrix}
\begin{bmatrix}
{\rm vec}\{\langle\bar{\sigma}\rangle_{-}\}\\
{\rm vec}\{\langle\bar{\sigma}\rangle_{+}\}\\
\end{bmatrix} 
= 
\begin{bmatrix}
{\rm vec}\{B_{-}\}\\
{\rm vec}\{B_{+}\}
\end{bmatrix},
\label{vec_sigmal2_matrix}
\end{equation}
and after the second vectorization as
\begin{equation}
{\rm vec}^2\{A\} {\rm vec}^2\{\bar{\sigma}\} = {\rm vec}^2\{B\},
\label{vec_sigmal2}
\end{equation}
where
\begin{eqnarray}
(A_{--})^{ml}_{nk} &=& \delta_{mn}\delta_{lk} - \langle\kappa\rangle_{nk} \Theta_{k-n+1} \delta_{nl},
\nonumber\\
(A_{-+})^{ml}_{nk} &=&  - \langle\bar{s}\rangle^{m}_{nk} \Theta_{n-k+1} \delta_{nl},
\nonumber\\
(A_{+-})^{ml}_{nk} &=& - \langle\bar{\kappa}\rangle^{m}_{nk} \Theta_{k-n+1} \delta_{nl},
\nonumber\\
(A_{++})^{ml}_{nk} &=& \delta_{mn}\delta_{lk} - \langle s\rangle_{nk} \Theta_{n-k+1} \delta_{nl},
\label{A2}
\end{eqnarray}
\begin{equation}
B_{+}^{ml} = B_{-}^{ml} = 1,
\label{D2}
\end{equation}
Finally, the criticality condition is
\begin{equation}
\det({\rm vec}^2\{A\}) = 0.
\label{critical2}
\end{equation}

\subsection{$L=1$}

When there is only one complex layer ($L(\chi)=1$) from Eq. (\ref{A2}) we obtain
\begin{equation}
\det({\rm vec}^2\{A\}) =
\begin{vmatrix}
1 & -\langle k\rangle_{11} & -\langle\bar{s}\rangle^0_{10} & -\langle s\rangle_{11}\\
0 & 1-\langle k\rangle_{11} & -\langle s\rangle_{10} & -\langle\bar{s}\rangle^1_{11}\\
-\langle\bar{k}\rangle^1_{01} & 0 & 1 & 0\\
0 & -\langle\bar{k}\rangle^1_{11} & -\langle s\rangle_{10} & 1-\langle s\rangle_{11}
\end{vmatrix}.
\label{detA2L1}
\end{equation}
One can go ahead and expand the determinant. However, I have not found any grouping of the terms that leads to a compact algebraic equation. Therefore, I'll work straight with Eq. (\ref{detA2L1}).

\begin{figure}[t]
\includegraphics[width=3.3in]{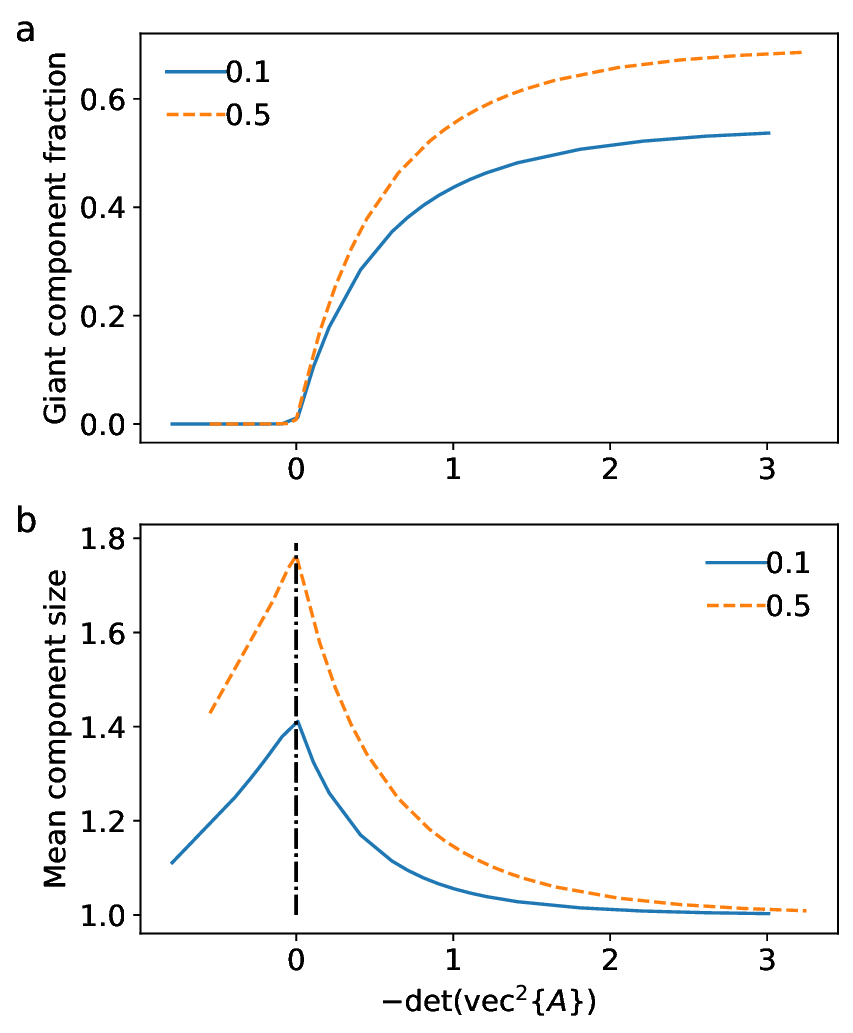}
\caption{{\bf With intra-layer inclusions.} Numerical estimation of (A) the giant component fraction and (B) the mean component size as a function of $-\det({\rm vec}\{ {\rm vec}\{A\}\})$ for a random chygraph with one layer of $n_0$ atoms, one layer of $n_1$ complexes, $m_0$ inclusions of a randomly selected atom into a randomly selected complex and $m_1$ inclusions of a randomly selected complex into a randomly selected complex. The model parameters are $n_0=n_1=10^6$, two values of $\theta_0=m_0/n_0$ as indicated in the legend and $\theta_2=m_1/n_1$ in the interval $[0.1,2]$. Average was taken over 10 random chygraphs.}
\label{fig_random_chygraph2}
\end{figure}

\subsection{Numerical example}

To test Eq. (\ref{detA2L1}) I consider a random chygraph with a layer of $n_0$ atoms, a layer of $n_1$ complexes, $m_0$ inclusions of a randomly selected atom into a randomly selected complex and $m_1$ inclusions of a randomly selected complex into a randomly selected complex. A possible realization of this model is illustrated in Fig. \ref{fig_publications}.

Since inclusions are random both the chydegrees and the component sizes have Poisson distributions with averages $\langle\kappa\rangle_{01}=m_0/n_0$, $\langle\kappa\rangle_{11} = m_1/n_1$, $\langle s\rangle_{10} = m_0/n_1$ and $\langle s\rangle_{11}=m_1/n_1$. For the Poisson distribution the excess average coincides with the average and therefore $\langle\bar{\kappa}\rangle^1_{01}=m_0/n_0$, $\langle\bar{\kappa}\rangle^1_{11} = m_1/n_1$, $\langle \bar{s}\rangle^0_{10} = m_0/n_1$ and $\langle \bar{s}\rangle^1_{11}=m_1/n_1$. Substituting these values into Eq. (\ref{detA2L1}) we obtain
\begin{equation}
\det({\rm vec}^2\{A\}) =
\begin{vmatrix}
1 & -\theta_1 & -r\theta_0 & -\theta_1\\
0 & 1-\theta_1 & -r\theta_0 & -\theta_1\\
-\theta_0 & 0 & 1 & 0\\
0 & -\theta_1 & -r\theta_0 & 1-\theta_1
\end{vmatrix},
\label{detA2L1_random}
\end{equation}
where $r=n_0/n_1$ and $\theta_i = m_i/n_i$, $i=0,1$.

Figure \ref{fig_random_chygraph2} reports the numerical estimation of the giant component fraction and the mean component size as a function of $-\det({\rm vec}^2\{A\})=0$, computed using Eq. (\ref{detA2L1_random}). There is a phase transition at $-\det({\rm vec}^2\{A\})=0$, with the emergence of a giant component and a maximum of the mean component size excluding the giant component. The agreement demonstrates the validity of the analytical results for chygraphs with intra-layer connections.

\section{Conclusions}
\label{conclusions}

In conclusion, chygraphs are a versatile combinatorial structure to represent complex systems. They allow for encoding different types of structural heterogeneities and hierarchical constructions. The key ingredient is the fractal nature of the chygraph: a complex is composed of atoms and other complexes and it can be part of other complexes as well. I have calculated the component sizes statistics of chygraphs using vectorization. Future work is required to extend this formalism to other problems.



%

\end{document}